\documentclass[twocolumn, aps, prb]{revtex4}
\usepackage{graphicx}
\usepackage{amssymb}
\usepackage{amsmath}

\newcommand{\GVec}[1]{\mbox{\boldmath$#1$}}

\def\Vec#1{{\bf #1}}
\def\GVec#1{\mbox{\boldmath $#1$}}

\def\H{{\mathcal H}}
\def\vare{\varepsilon}
\def\av#1{\langle #1 \rangle}

\def\partd#1#2{\frac{\partial #1}{\partial #2}}
\def\partdd#1#2{\frac{\partial^2 #1}{\partial #2^2}}

\begin{document}

\title{Diamagnetism in disordered graphene}
\author{Mikito Koshino and Tsuneya Ando}
\affiliation{
Department of Physics, Tokyo Institute of Technology
2-12-1 Ookayama, Meguro-ku, Tokyo 152-8551, Japan}
\date{\today}

\begin{abstract}
The orbital magnetism is studied in graphene monolayer within the
effective mass approximation.  In models of short-range and long-range
disorder, the magnetization is calculated with self-consistent Born
approximation.  In the zero-field limit, the susceptibility becomes
highly diamagnetic around zero energy, while it has a long tail
proportional to the inverse of the Fermi energy.  We demonstrated how
the magnetic oscillation vanishes and converges to the susceptibility,
on going from a strong-field regime to zero-field. 
The behavior at zero energy is shown to be highly singular.
\end{abstract}

\maketitle

\section{Introduction}

The monolayer graphene has a band structure 
analogous to the massless relativistic particle, and 
its peculiar electronic properties have attracted 
much interest. 
Recently several experimental techniques
make atomically thin graphene sheets available,\cite{Novo04,Berg,Novo05,Zhan05-2}
and the nature of this unique system is being revealed.
In this paper we present a theoretical study on the orbital magnetism of
graphene including the disorder effects.

The graphene has a semi-metallic electronic structure where
the conduction and valence bands touch 
at the Brillouin zone corners, $K$ and $K'$ points.
Around the band touching point (set to $\vare =0$),
the low energy spectrum has a linear dispersion 
analogous to the massless Dirac Fermion.
The spectrum in a magnetic field
is different from that in usual metals
in that the Landau level spacing is not even 
but wider in lower energies, and is proportional to $\sqrt{B}$,
not to $B$, where $B$ is the magnetic field, \cite{McCl}
and this leads to an unusual behavior in the orbital magnetization.
The magnetism of graphene was first studied 
as a simple model for three-dimensional (3D) graphite,\cite{McCl}
where the susceptibility of
the disorder-free graphene was calculated within the effective mass
approximation. It was found that the system exhibits
a large diamagnetism at $\vare_F =0$,
expressed as a delta function of $\vare_F$
at the absolute zero temperature.
The graphene magnetism was considered again
in studies on the graphite intercalation compounds,
where the tight-binding model was applied
for a wide range of Fermi energies.\cite{Shar,Safr,Blin,Sait}

In the presence of the disorder,
it becomes nontrivial how the magnetization behaves
under this unusual electronic structure.
Particularly, it is not clear how the delta-function in 
the susceptibility is broadened,
since we naively suppose that the scattering is absent
at $\vare=0$ where the density of states vanishes.
Moreover, we do not know how the magnetic oscillation is destroyed
by the disorder when we go from the high-field to the low-field regime,
and how it converges to the zero-field limit.

The effects of disorder on graphene under magnetic fields 
have been examined in early theoretical studies
before the experimental discovery of graphene,
where the electronic structure,\cite{Shon} 
the transport properties,\cite{Shon,Zhen,Ando02}
and the de Haas-van Alphen effect\cite{Gusy04} were investigated.
More recently the Shubnikov-de Haas oscillation was studied
in disordered graphene,\cite{Gusy05,Gusy06}
and the spectral and transport properties were examined 
in presence of lattice defects under magnetic fields.\cite{Peres}

The purpose of this paper is to calculate the magnetization
of disordered graphene in arbitrary magnetic fields,
and to obtain the perspective which connects the
high-field and zero-field limit.
For the model disorder,
we introduce the short-ranged and long-ranged scatterers
following the formulation in Ref.\ \onlinecite{Shon,Zhen,Ando02},
and treat the disorder effects within 
a self-consistent Born approximation (SCBA).
The paper is organized as follows:
In Sec.\ \ref{sec_form} we briefly discuss the effective mass
Hamiltonian and the SCBA 
in order to make this paper self-contained although 
fully discussed previously.\cite{Shon}
The analytic discussions of the magnetization 
in the zero-field limit 
and the numerical calculation for finite fields
are presented in Sec.\ \ref{sec_mag}.
Discussions and conclusions are given in Sec.\ \ref{sec_disc}.

\section{Formulation}
\label{sec_form}
\subsection{Hamiltonian}

We start with the effective mass Hamiltonian 
in an ideal graphene in a magnetic field given by \cite{Shon}
\begin{equation}
 \H_0 = \frac{\gamma}{\hbar}
\begin{pmatrix}
 0 & \pi_x - i \pi_y & 0 & 0 \\
 \pi_x + i \pi_y & 0 & 0 & 0 \\
 0 & 0 & 0 & \pi_x + i \pi_y \\
 0 & 0 & \pi_x - i \pi_y & 0
\end{pmatrix},
\label{eq_H}
\end{equation}
where $\GVec{\pi} = \Vec{p} + e\Vec{A}$ with
the electron momentum operator $\Vec{p}$ and the vector
potential $\Vec{A}=(0,Bx)$ in the Landau gauge,
and $\gamma = \sqrt{3}a \gamma_0 /2$ with $a$ being the lattice constant 
and $\gamma_0$ the hopping integral between nearest-neighbor
carbon atoms.
A graphene is composed of a honeycomb network of carbon atoms,
where a unit cell contains a pair of sublattices, denoted by $A$ and $B$.
The Hamiltonian (\ref{eq_H}) operates on a four-components wave function
$(F^K_A,F^K_B,F^{K'}_A,F^{K'}_B)$, where $F^K_A$ and $F^K_B$
represent the envelope functions at $A$ and $B$ sites for $K$ point,
respectively, and $F^{K'}_A$ and $F^{K'}_B$ for $K'$.

The eigenstates are labeled by $(j,n,k)$
with the valley index $j=K,K'$, the Landau level index
$n = 0,\pm 1, \dots$, and the wave vector $k$ along
$y$ direction.\cite{Shon}
The eigenenergy depends solely on $n$ as
$\varepsilon_n = \hbar\omega_B\,\,{\rm sgn}(n)\sqrt{|n|}$,
where $\hbar\omega_B = \sqrt{2}\gamma/l$ with
$l = \sqrt{\hbar/eB}$.
The wave functions are written as
\begin{eqnarray}
 \Vec{F}^K_{nk} &=& \frac{C_n}{\sqrt{L}}\exp(iky)
\left(
\begin{array}{c}
{\rm sgn}(n) (-i) \phi_{|n|-1,k}\\
\phi_{|n|,k}\\0\\0
\end{array}
\right), \\
 \Vec{F}^{K'}_{nk} &=& \frac{C_n}{\sqrt{L}}\exp(iky)
\left(
\begin{array}{c}
0\\0\\
\phi_{|n|,k}\\
{\rm sgn}(n) (-i) \phi_{|n|-1,k}
\end{array}
\right),
\label{eq_LL}
\end{eqnarray}
where
$\phi_{n,k}(x) = (2^{n}n!\sqrt{\pi}l)^{-1/2} \,\,e^{-z^2/2}H_{n}(z)$,
with $z = (x+kl^2)/l$ and $H_n$ being the Hermite
polynomial, and 
\begin{eqnarray}
 C_n = \left\{
\begin{array}{cc}
 1 & (n=0), \\
 1/\sqrt{2} & (n\neq 0),
\end{array}
\right.
\nonumber\\
 {\rm sgn}(n) = \left\{
\begin{array}{cc}
 0 & (n=0), \\
 n/|n| & (n\neq 0).
\end{array}
\right.
\end{eqnarray}

For the disorder potential, 
we consider two simple models:
short- and long-ranged scatterers.\cite{Shon}
The first is on-site potential
localized at a particular $A$ or $B$ sites with a random amplitude.
A scatterer on $A$ site at $\Vec{R}_A$ 
is represented as 
\begin{equation}
 U(\Vec{r}) = 
\left(
\begin{array}{cccc}
 1 & 0 & z_A^* z'_A & 0 \\
 0 & 0 & 0 & 0 \\
z_A {z'_A}^* & 0 & 1 & 0 \\
 0 & 0 & 0 & 0
\end{array}
\right)
u_i \delta(\Vec{r}-\Vec{R}_A),
\end{equation}
and that for $B$ site at $\Vec{R}_B$ as
\begin{equation}
 U(\Vec{r}) = 
\left(
\begin{array}{cccc}
 0 & 0 & 0 & 0 \\
0 & 1 & 0& z_B^* z'_B \\
 0 & 0 & 0 & 0 \\
0 & z_B {z'_B}^*  & 0 & 1 
\end{array}
\right)
u_i  \delta(\Vec{r}-\Vec{R}_B),
\end{equation}
where we introduced
$z_X = e^{i\Vec{K}\cdot\Vec{R}_{X}}$, $z'_X = e^{i\Vec{K}'\cdot\Vec{R}_{X}}$
with $X = A$ and $B$, and $u_i = (\sqrt{3}a^2/2)U_i$
with the on-site energy $U_i$.
We assume that the scatterers are equally distributed
on $A$ and $B$ sites with density $n_i^A = n_i^B = n_i/2$ and the
mean square amplitude $\av{(u_i^A)^2}=\av{(u_i^B)^2}=u_i^2$.

Dominant scatterers in graphenes are expected to have their potential
range larger than the lattice constant for which inter-valley scattering
is much smaller than intra-valley scattering.  Further, realistic
scatterers are likely to have the range comparable to the Fermi
wavelength.\cite{Nomu06,Ando06,Nomu07} In the following, however, we shall
assume scatterers with potential range smaller than the Fermi
wavelength.  The reason is that the results are expected to remain
qualitatively the same and further that actual calculations are
practically possible.

In this long-range model, a scatterer at $\Vec{R}$ is expressed by
\begin{equation}
 U(\Vec{r}) = 
\left(
\begin{array}{cccc}
 1 & 0 & 0 & 0 \\
0 & 1 & 0& 0 \\
 0 & 0 & 1 & 0 \\
0 &  0  & 0 & 1 
\end{array}
\right)
u_i  \delta(\Vec{r}-\Vec{R}).
\end{equation}
We assume the scatterer density $n_i$ and the mean square amplitude $u_i^2$.
It was shown that the transport properties in the short-ranged disorder
and the long-ranged one are qualitatively similar.\cite{Shon,Zhen,Ando02}

\subsection{Self-Consistent Born Approximation (SCBA)}

We introduce the self-consistent Born approximation
for graphene, following the formulation in Ref.\ \onlinecite{Shon}.
The self-energy of the
disorder-averaged Green's function $\langle G_{\alpha,\alpha'}\rangle$
is given by
\begin{equation}
\Sigma_{\alpha,\alpha'} (\vare) = 
\sum_{\alpha_1, \alpha_1'}
\langle 
U_{\alpha,\alpha_1}U_{\alpha'_1,\alpha'}
\rangle
\langle 
G_{\alpha_1,\alpha'_1}(\vare)
\rangle,
\label{eq_SCBA1}
\end{equation}
with $\alpha= (j,n,k)$, where $\av{\cdots}$ represents the average over the impurity configurations.

In the short-range model, the self-energy and thus
the averaged Green's function become diagonal with respect to $\alpha$,
and further, the self-energy is independent of $\alpha$.
We then have
\begin{eqnarray}
&& \av{G_{\alpha,\alpha '}(\vare)} = \delta_{\alpha,\alpha '} G_\alpha(\vare) ,
\\
&& G_\alpha(\vare) = \frac{1}{\vare - \vare_\alpha- \Sigma(\vare)} ,
\end{eqnarray}
where $\Sigma(\varepsilon)$ is the self-energy.
The self-consistent equation (\ref{eq_SCBA1}) is explicitly written as
\begin{eqnarray}
\Sigma (\vare)= \frac{W(\hbar\omega_B)^2}{2}
\sum_{n=-\infty}^{\infty} \frac{g(\vare_n)}{\vare-\vare_n-\Sigma(\vare)},
\label{eq_SCBA}
\end{eqnarray}
where we introduced a cutoff function $g(\vare)$ which 
is 1 in $|\vare| \ll \vare_c$ and smoothly
vanishes around $\vare = \pm\vare_c$. For example we can take
$g(\vare) = \vare_c^\alpha/(|\vare|^\alpha + \vare_c^\alpha)$
with $\alpha \geq 2$.
Further, $W$ is the dimensionless parameter for the disorder strength
defined as
\begin{equation}
 W = \frac{n_i u_i^2}{4\pi\gamma^2}.
\label{eq_w}
\end{equation}
The density of states per a unit area is given by
\begin{equation}
 \rho(\vare) 
= -\frac{g_v g_s}{2\pi^2 \gamma^2 W}{\rm Im}\Sigma(\vare+i0),
\label{eq_dos}
\end{equation}
where $g_v=g_s=2$ is the valley and spin degeneracy, respectively.

In the zero-field limit, (\ref{eq_SCBA}) becomes 
\begin{eqnarray}
 \Sigma (\vare)
= 2W \int_0^\infty tdt
\frac{(\vare-\Sigma)g(t)}{(\vare-\Sigma)^2-t^2}.
\label{eq_sigma0}
\end{eqnarray}
The integral is approximately written in $\vare \ll \vare_c$ as
\begin{eqnarray}
  \Sigma(\vare) = -W (\vare-\Sigma) \log 
\left(-\frac{\vare_c^2}{(\vare-\Sigma)^2}\right),
\end{eqnarray}
where the branch of $\log$ must be appropriately chosen.
Then we can solve this equation analytically,
\begin{eqnarray}
  \Sigma (\vare) = \vare - 
\vare\left[2Wf_L \left(-\frac{i\vare}{2W\Gamma_0} \right)
\right]^{-1}.
\label{eq_sigma_anl}
\end{eqnarray}
where $f_L(z)$ is the Lambert W-function,
which is defined as the inverse function of  $z = y e^y$,
and
\begin{equation}
\Gamma_0 = \vare_c \exp\Big(- {1 \over 2W} \Big) .
\label{eq_g0}
\end{equation}
At $\vare = 0$ in particular, we have
\begin{equation}
\Sigma(0+i0) = -i \Gamma_0.
\end{equation}
In $|\vare| \gg \Gamma_0$,  $\Sigma$ is approximately written 
with use of the expansion $f_L(z) \approx \log(z) - \log\log(z)$
for $|z| \gg 1$ as, 
\begin{eqnarray}
 \Sigma (\vare+i0)
\approx -2W \vare 
\log \left|\frac{\vare_c}{\vare}\right| - i \pi W|\vare|.
\label{eq_sigma_boltz}
\end{eqnarray}
This can be alternatively derived from
(\ref{eq_sigma0}) with assuming $|\vare| \gg |\Sigma|$,
and thus corresponds to the Boltzmann limit.
If $W \sim 1$, the states around $\vare=\vare_c$
are completely mixed up with those at $\vare=0$,
as expected from the imaginary part of $\Sigma$ in (\ref{eq_sigma_boltz}).
To avoid this undesirable situation, we assume $W \ll 1$
in the following calculation.

When the magnetic field is large enough that
a Landau level is well separated from others,
(\ref{eq_SCBA}) can be approximately solved around the energy
of that level.
The width of the Landau level is estimated as $2\Delta$ with
\begin{equation}
\Delta = \sqrt{2W}\hbar\omega_B.
\label{eq_gmom}
\end{equation}

In the long-ranged model, the self-energy and Green's function  
have off-diagonal matrix elements between $(j,n,k)$
and $(j,-n,k)$. We have
\begin{eqnarray}
\Sigma_{\alpha,\alpha '}(\vare) &=& 
\delta_{j,j'}\delta_{k,k'} [\delta_{n,n'}\Sigma^{\rm d}(\vare) +
\delta_{n,-n'}\Sigma^{\rm o}(\vare)] , 
\end{eqnarray}
Introducing 
$\Sigma^\pm \equiv \Sigma^{\rm d} \pm \Sigma^{\rm o}$,
the equation becomes
\begin{eqnarray}
\Sigma^+ (\vare) &=& 
W(\hbar\omega_B)^2
\sum_{n=0}^{\infty} 
\frac{(\vare-\Sigma^-)g(\vare_n)}
{(\vare-\Sigma^+)(\vare-\Sigma^-)-\vare_n^2} , \quad
\label{eq_SCBA_long1}
\\
\Sigma^- (\vare) &=& 
W(\hbar\omega_B)^2
\sum_{n=1}^{\infty} 
\frac{(\vare-\Sigma^+)g(\vare_n)}
{(\vare-\Sigma^+)(\vare-\Sigma^-)-\vare_n^2},
\label{eq_SCBA_long2}
\end{eqnarray}
with the same $W$ as the short-range case Eq.\ (\ref{eq_w}). 
The density of states per a unit area becomes
\begin{equation}
 \rho(\vare) = 
 -\frac{g_v g_s}{2\pi^2 \gamma^2 W}\frac{1}{2}{\rm Im}
\left[\Sigma^+(\vare+i0) + \Sigma^-(\vare+i0)
\right].
\label{eq_dos_long}
\end{equation}

In a high magnetic field such that Landau levels are well separated,
the width of the Landau level becomes the same as
$\Delta$ in (\ref{eq_gmom}) for the level $N \neq 0$,
while it is $\sqrt{2}\Delta$ for $N=0$.
In the weak-field limit,  $\Sigma^+$ and $\Sigma^-$ coincide and 
satisfy (\ref{eq_sigma0}).

\subsection{Magnetization and susceptibility}

The magnetization is defined as
\begin{equation}
 M = -\left(\frac{\partial \Omega}{\partial B}\right)_{\mu} ,
\end{equation}
where $\Omega(T,\mu,B)$ is the thermodynamic potential and
$\mu$ is the chemical potential.
By noting that the electron concentration $N$ is given by
\begin{equation}
 N = -\left(\frac{\partial \Omega}{\partial \mu}\right)_{B},
\end{equation}
we obtain so-called Maxwell's relation,
\begin{equation}
 \left(\frac{\partial M}{\partial \mu}\right)_{B}
= \left(\frac{\partial N}{\partial B}\right)_{\mu}.
\label{eq_dmdmu}
\end{equation}
We write $N$ in terms of the density of states $\rho$ as
\begin{equation}
N = \int_{-\infty}^{\infty} 
\rho(\vare,B)f(\vare)d\vare,
\end{equation}
with $f(\vare)=1/(1+e^{(\vare-\mu)/k_B T})$,
and calculate $M$ by integrating
(\ref{eq_dmdmu}) over $\mu$.
After a little algebra, we obtain
\begin{eqnarray}
 M = 
\int_{-\infty}^{\infty} d\vare f(\vare)
\int_{-\infty}^\vare d\vare'
\frac{\partial \rho(\vare',B)}{\partial B}.
\label{eq_m}
\end{eqnarray}
In SCBA, we evaluate this by substituting $\rho$ 
with (\ref{eq_dos}) or (\ref{eq_dos_long})
depending on the type of the disorder.
The magnetization in a nonzero temperature is always written
in terms of that of $T=0$ as
\begin{equation}
 M(T,\mu) = \int_{-\infty}^\infty d\vare 
\left(-\frac{\partial f(\vare)}{\partial \vare}\right)
M(0, \vare).
\label{eq_finite_temp}
\end{equation}
The magnetic susceptibility is given by
\begin{equation}
\chi = \left.\frac{\partial M}{\partial B} \right|_{B= 0} ,
\end{equation}
taking the zero-field limit in (\ref{eq_m}).

\section{Magnetization in disordered graphenes}
\label{sec_mag}


For the short-ranged disorder,
Eqs.\ (\ref{eq_SCBA}), (\ref{eq_dos}), and (\ref{eq_m})
lead to the expression
\begin{eqnarray}
\chi =
-\frac{g_vg_s}{6\pi^2}\frac{e^2\gamma^2}{\hbar^2}\,\,
\int_{-\infty}^\infty 
d\vare f(\vare)
{\rm Im} \frac{1}{(\vare-\Sigma(\vare))^2},
\label{eq_chi}
\end{eqnarray}
where $\Sigma$ is the self-energy at $B=0$.
The derivation of this is straightforward
and is presented in Appendix \ref{sec_appe}.

In the energy range $\vare \ll \vare_c$, we can use the explicit form
(\ref{eq_sigma_anl}) for $\Sigma$.
At absolute zero temperature, we execute the integral to have
\begin{eqnarray}
 \chi(\vare_F) =
-\frac{g_vg_s}{3\pi^2}\frac{e^2\gamma^2}{\hbar^2}\,\,
\frac{2W}{\Gamma_0} F \left(\frac{\vare_F}{2W\Gamma_0} \right),
\label{eq_chi_anl}
\end{eqnarray}
with
\begin{eqnarray}
 F(x) = -\frac{1}{x}{\rm Im} \left[
f_L(-ix) + \frac{1}{2}f_L^2(-ix)
\right].
\end{eqnarray}
The function $F(x)$ has the maximum at $x=0$ with $F(0) = 1$,
giving
\begin{equation}
 \chi(0) =
-\frac{g_vg_s}{3\pi^2}\frac{e^2\gamma^2}{\hbar^2}\,\,
\frac{2W}{\Gamma_0}.
\label{eq_peak}
\end{equation}
In the energy range $|\vare| \gg \Gamma_0$, we use (\ref{eq_sigma_boltz}) and obtain
\begin{eqnarray}
\chi(\vare_F)
\approx
-\frac{g_vg_s}{3\pi}\frac{e^2\gamma^2}{\hbar^2}\,\,
\frac{W}{|\vare_F|},
\label{eq_chi_boltz}
\end{eqnarray}
which monotonically decreases as $|\vare_F|$ increases.
The behavior of $\chi(\vare_F)$
can be roughly described as a long-tailed peak (\ref{eq_chi_boltz})
which saturates around $\vare \sim \Gamma_0$ 
to the value (\ref{eq_peak}).

When the disorder $W$ becomes smaller,
the peak of the susceptibility (\ref{eq_chi_anl})
becomes narrower and higher
as $\Gamma_0$ behaves as $\propto\exp(-1/2W)$.
The integral over $\vare$ rigorously becomes 
$-g_vg_se^2\gamma^2/(6\pi\hbar^2)$, as proved in Appendix \ref{sec_appe}.
This is roughly verified by integrating (\ref{eq_chi_boltz})
from $-\vare_c$ to $\vare_c$
with the region $|\vare| < \Gamma_0$ excluded and by using Eq.\ (\ref{eq_g0}).
Thus, in the clean limit $W\rightarrow 0$,
$\chi$ becomes a delta function,
\begin{eqnarray}
\chi(\vare_F) = 
-\frac{g_vg_s}{6\pi}\frac{e^2\gamma^2}{\hbar^2}\,\,
\delta(\vare_F),
\label{eq_chi_delta}
\end{eqnarray}
which agrees with the result in Refs.\ \onlinecite{McCl} and \onlinecite{Safr}.

\begin{figure}
\begin{center}
 \leavevmode\includegraphics[width=80mm]{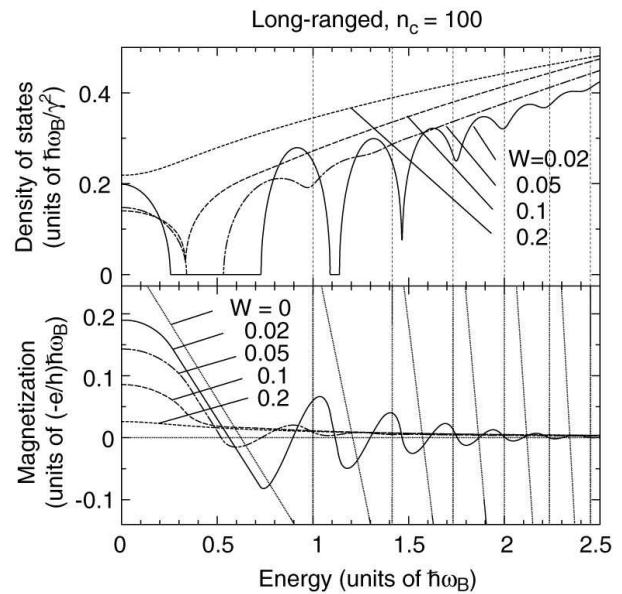}
\end{center}
\caption{
Density of states (above) and the magnetization 
at $T=0$ (below) in the long-ranged disorder 
with several strength $W$'s.
The plot is against the Fermi energy,
and the values are per a spin and per a valley.
Vertical dashed lines shows the energies of the Landau level
in the clean limit.
}
\label{fig_short_wdep}
\end{figure}


For the long-ranged disorder,
the expression for the susceptibility
in $|\vare| \ll \vare_c$ becomes,
\begin{eqnarray}
 \chi &\approx& 
-\frac{g_vg_s}{6\pi^2}
\frac{e^2\gamma^2}{\hbar^2}
\int_{-\infty}^{\infty} d\vare f(\vare)
\nonumber\\
&&\hspace{-5mm}
\times
{\rm Im}
\frac 1 {X^2}
\left[
1 -  \frac{3W}{1+ W\log(-\vare_c^2/X^2)}
\right],
\label{eq_chi_long}
\end{eqnarray}
with $X = \vare - \Sigma(\vare)$.
The derivation is given in Appendix \ref{sec_appe}.
Compared with the short-ranged case (\ref{eq_chi}),
we have the extra second term of the order of $O(W)$,
but this gives a minor effect since $W$ is assumed to be small.
When $O(W^2)$ is neglected, the susceptibility becomes
just $1-3W$ times as large as in the short-ranged disorder.
Accordingly the integration of $\chi$ over $\vare$ weakly depends on $W$,
while in $W\rightarrow 0$ we again get (\ref{eq_chi_delta}).


In a strong magnetic field where the Landau levels are resolved,
the magnetization exhibits an oscillatory behavior 
as a function of the Fermi energy and the magnetic field.
The damping of the magnetic oscillation in the disorder was discussed
in a simple approximation 
where the scattering rate was assumed to be constant.\cite{Gusy04}
We calculate here the magnetization at nonzero fields 
in SCBA, since this kind of treatment is essential 
in investigating the behavior at the zero energy.
We numerically evaluate (\ref{eq_m}),
in which the derivative in $B$ is taken with a finite increment $\Delta B$.
Here and the following 
we take the long-ranged disorder,
and plot every quantity per a spin and a valley.
The field amplitude $B$ is specified by 
$n_c = (\vare_c/\hbar\omega_B)^2 \propto 1/B$,
which represents how many Landau levels are accommodated between 
$\vare =0$ and $\vare_c$.
We set $n_c =100$ here.

As an overview of the dependence on the disorder strength,
we plot in Fig.\ \ref{fig_short_wdep} the density of states and 
the magnetization for 
several $W$'s
at zero temperature and a fixed magnetic field. 
The density of states is basically equivalent to that already obtained in Ref.\ \onlinecite{Shon},
but we present this here to demonstrate 
the relation to the magnetization.
We see that the Landau levels are separated more clearly
in the lower energy due to the larger level splitting,
and the magnetization exhibits an oscillation
in the corresponding region.
As $W$ becomes larger, the oscillatory part vanishes
from the higher-energy side.
The results for the short-range disorder are not shown, but 
qualitatively similar to those for the long-range disorder.

The Landau level broadening in disordered graphene 
has also been studied for a system with lattice vacancies.\cite{Peres}
The result becomes somewhat different from our model
in that the Landau levels around $\vare=0$ become much broader
than higher levels, in accordance with the fact that
the vacancies give rise to impurity states around 
the band touching point.\cite{Peres, Igam, Ando_et_al_1999b, Pereira}
We do not have a strong scattering enhancement at $\vare=0$
in the present effective-mass model,
where the on-site energy of the disorder potential
is assumed to be much smaller than the $\pi$-band width.

\begin{figure}
\begin{center}
 \leavevmode\includegraphics[width=80mm]{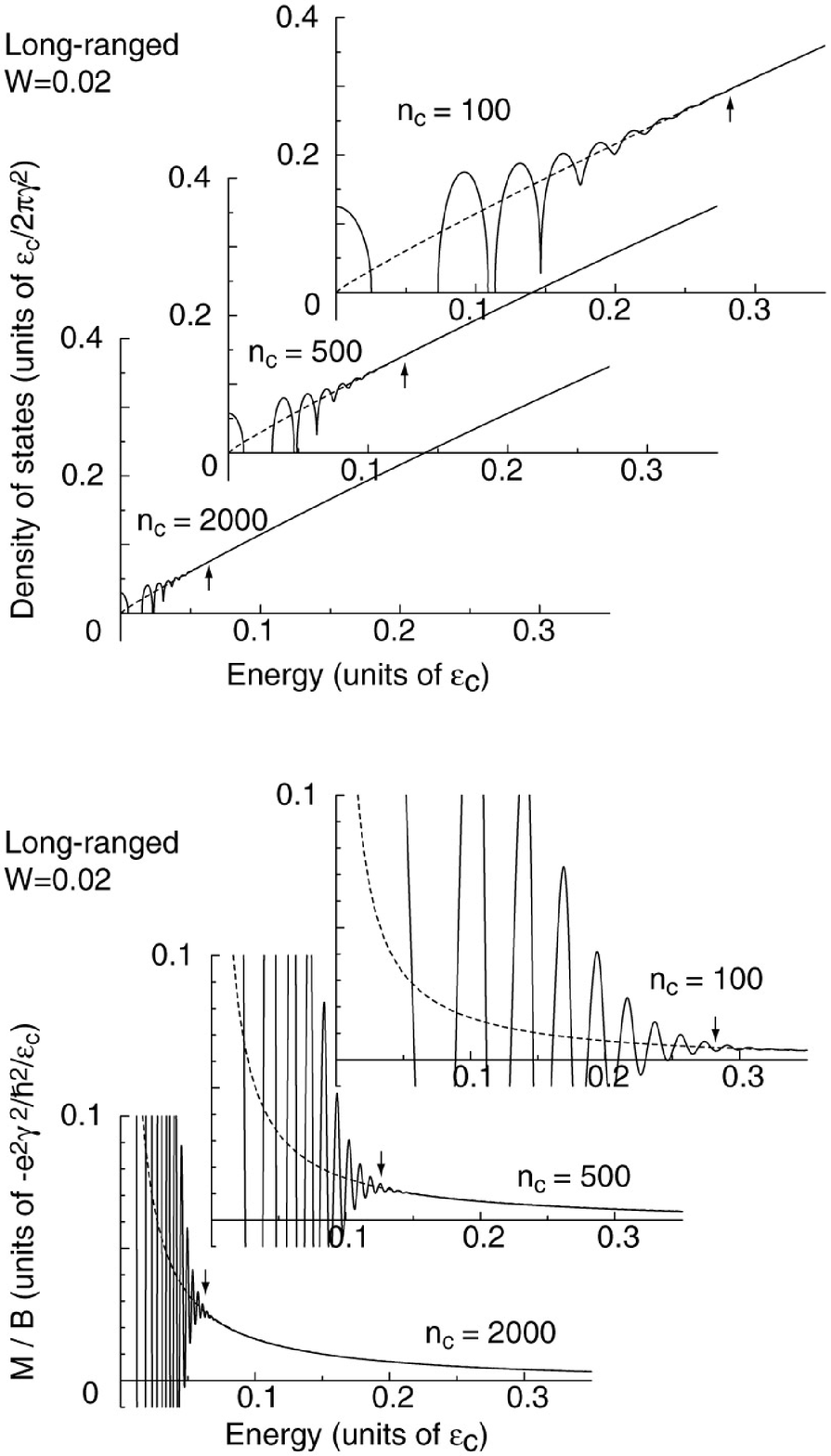}
\end{center}
\caption{
Density of states (above) and the magnetization
at $T=0$ (below) in the long-ranged disorder with $W=0.02$,
in several magnetic fields specified by $n_c = (\vare_c/\hbar\omega_B)^2$.
Dashed curves show the zero-field limit.
}
\label{fig_short_w002}
\end{figure}

We focus on the case of $W=0.02$ and 
show in Fig.\ \ref{fig_short_w002} 
the plots of the density of states and of $M/B$ for several 
different magnetic fields.
We see that the oscillation in $M$ terminates at a certain point
and in higher $\vare$
$M/B$ sticks to the zero-field limit $\chi$ shown as a dashed line.
The oscillation is observable when the Landau level spacing 
$\hbar\omega_B|\sqrt{n+1}-\sqrt{n}| \sim (\hbar\omega_B)^2/(2|\vare|)$
is larger than the energy 
broadening at $B=0$,
which is $\pi W|\vare|$ in the Boltzmann limit (\ref{eq_sigma_boltz}).
Then the condition becomes
\begin{equation}
\vare > \frac{\hbar\omega_B}{\sqrt{2\pi W}}.
\end{equation}
In Fig.\ \ref{fig_short_w002}, the boundary is indicated by an arrow,
which actually divides the oscillating and non-oscillating parts.

\begin{figure}
\begin{center}
 \leavevmode\includegraphics[width=70mm]{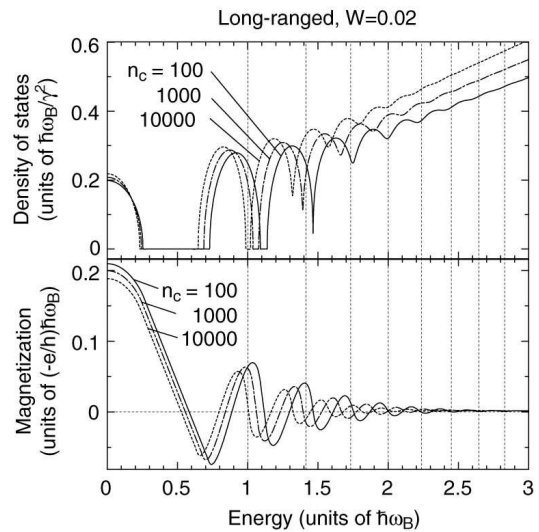}
\end{center}
\caption{
Density of states (above) and the magnetization 
at $T=0$ (below) in the long-ranged disorder with $W=0.02$,
in several magnetic fields.
The vertical and horizontal
axes are renormalized in units of factors $\propto \hbar\omega_B$.
Dashed vertical lines are the energies of the Landau levels
in the ideal limit.}
\label{fig_short_w002R}
\end{figure}

We show in Fig.\ \ref{fig_short_w002R} the renormalized
density of states and magnetization against $\vare/\hbar\omega_B$.
We can see that the Landau level width
is almost independent of the magnetic field in this scale,
as expected from (\ref{eq_gmom}) in the strong-field limit,
while each level shifts toward zero energy
as $B$ becomes smaller ($n_c$ larger).
From the real part of $\Sigma$ in (\ref{eq_sigma_boltz}),
the shift can be estimated as $\Delta \vare_n \sim - W \vare_n \log |n_c/n|$.
The amplitude of the magnetic oscillation
roughly scales as $M \propto \hbar\omega_B \propto \sqrt{B}$,
in contrast to the behavior in the non-oscillating region 
where the relation $M = \chi B$ is valid. 
This is because the gain of the total energy $U$ due to the magnetic field
is proportional both to the Landau level spacing  $(\propto \sqrt{B})$
and the level degeneracy $(\propto B)$,
which gives $M \sim -dU/dB \propto \sqrt{B}$.\cite{McCl}
The oscillation amplitude gradually reduces as $B$ becomes smaller,
as the level shift causes reduction of the energy gap.

\begin{figure}
\begin{center}
 \leavevmode\includegraphics[width=80mm]{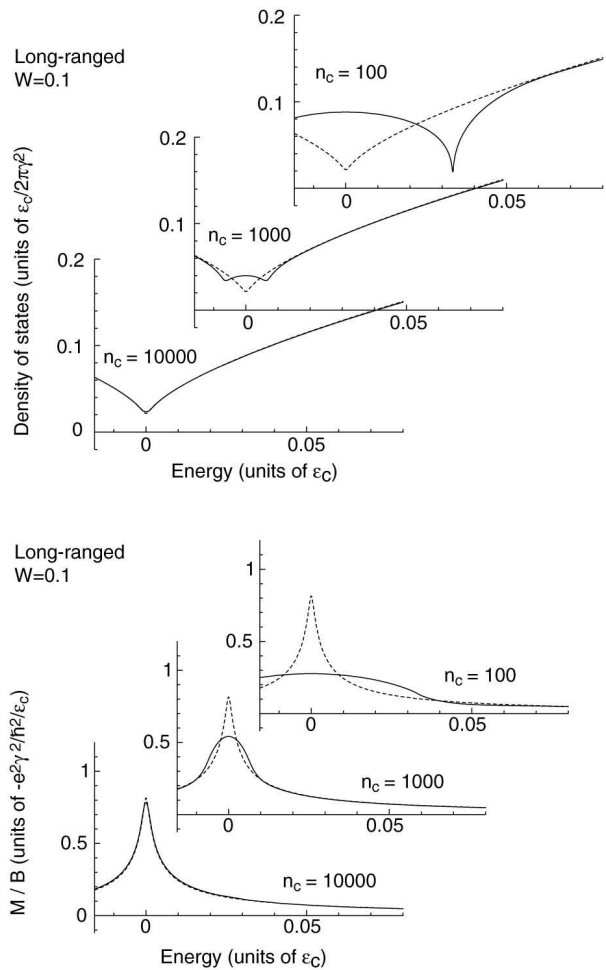}
\end{center}
\caption{
Density of states (above) and the magnetization 
at $T=0$ (below)
in the long-ranged disorder with $W=0.1$,
in several magnetic fields.
Dashed curves show the zero-field limit.}
\label{fig_short_w01}
\end{figure}

We expect that the lowest Landau gap vanishes
when the gap width is as small as the energy broadening
at $\vare =0$ in zero field, or
\begin{equation}
 \hbar\omega_B \sim \Gamma_0.
\label{eq_cond}
\end{equation} 
This is equivalent with the condition
that the first Landau level is shifted onto $\vare=0$,
or $|\Delta \vare_1| \sim \hbar\omega_B$, as naturally expected.
To focus on this critical behavior, 
we present in Fig, \ref{fig_short_w01} 
the plots of the density of states and the magnetization for $W=0.1$, 
where the condition (\ref{eq_cond}) is achieved at $n_c \sim 20000 $.
We see that the structure of the Landau level $n=0$ still survives at
$n_c = 100$, and $M/B$ deviates largely from $\chi$ around this region.
Here the magnetization at $\vare=0$ roughly scales as $M \propto \sqrt{B}$.
At $n_c = 10000$, the gap collapses 
and the magnetization peak almost reaches that of $\chi$,
and thus $M \propto B$.
If we fix the Fermi energy at $\vare=0$
and start the magnetic field from 0,
the magnetization should exhibit the crossover from
linear to square behavior in $B$.

\section{Discussion and Conclusion}
\label{sec_disc}

At nonzero temperatures, the magnetization 
as a function of the Fermi energy, $M(\vare)$, is smoothed
in accordance with (\ref{eq_finite_temp}), so that
fine structures smaller than $k_BT$ are smeared out.
This effect competes with 
energy broadening due to the impurity scattering, denoted here as $\Gamma$.
We expect the crossover from the high-field (magnetic oscillation)
to low-field regime ($M = \chi B$) occurs 
when either of $k_B T$ or $\Gamma$
exceeds the Landau level spacing $\Delta\vare$.
In a usual 2D metal with a constant level spacing,
it is known that the disorder effects
can be effectively included as the Dingle temperature, 
$k_B T_D = \Gamma/\pi$.
The reduction of the magnetic oscillation in disordered graphenes 
was studied with a constant $\Gamma$
and discussed with respect to the Dingle temperature \cite{Gusy04}.

In the realistic samples used in the experiment, dominant scatterers
are supposed to be screened charged impurities.\cite{Nomu06,Ando06,Nomu07}
There the scattering matrix elements 
between the states on a Fermi surface
are proportional to $1/k_F$, not a constant
like in our simple model. This situation
is effectively modeled in our calculation 
by assuming that the parameter
$W$ depends on $\vare_F$ as $W \propto 1/\vare_F^2$
in the long-range model.
Then we expect that the susceptibility in the Boltzmann limit 
(\ref{eq_chi_boltz}) becomes $\chi \propto 1/\vare_F^3$.
From the experimental value of the mobility of monolayer graphene,
we estimate $W \sim 70/\vare_F^2$ where $\vare_F$ is measured in units of meV.

The magnetization becomes highly singular
at zero energy in our model,
because the energy broadening, $\Gamma_0$, 
is exponentially small here.
In the case of charged impurities, however,
the scattering rate at zero energy may not be small
since the screening effect is strongly suppressed
due to the lack of the density of states.\cite{Ando06,Nomu07}
We need a self-consistent calculation 
including the screening and the disorder to study such a case.
This is out of the scope of this paper and left for a future study.

The experimental measurements of the magnetization 
of two-dimensional electron systems were performed 
on the semiconductor heterostructures,
by using the superconducting quantum interference device (SQUID)\cite{Stor,Mein}
or using the torque magnetometer.\cite{Eise,Pott,Wieg,Zhu}
We expect that the detection of the graphene magnetism
is also feasible with those techniques.

To summarize, we have studied the magnetization in graphene monolayer
in presence of the disorder with the effective mass model
and the self-consistent Born approximation.
The susceptibility $\chi(\vare_F)$ 
has a sharp diamagnetic peak around zero energy even in the disorder,
and a long tail proportional to the inverse of the Fermi energy.
We have demonstrated that
with the decrease of the magnetic field, 
the magnetic oscillation vanishes, and $M/B$ converges to $\chi$
as the Landau gaps are smeared out.

\section*{ACKNOWLEDGMENTS}

This work has been supported in part by the 21st Century COE Program at
Tokyo Tech \lq\lq Nanometer-Scale Quantum Physics'' and by Grants-in-Aid
for Scientific Research from the Ministry of Education, 
Culture, Sports, Science and Technology, Japan.

{\it Note added in proof:} --- After completion of this work,
we became aware of related work, Ref.\ \onlinecite{Fuku07}.


\appendix

\section{Susceptibility}
\label{sec_appe}


We present here the derivation of
the susceptibility in SCBA for the short-ranged disorder
Eq.\ (\ref{eq_chi}) and for the long-ranged (\ref{eq_chi_long}).
For the short-ranged case,
we obtain from (\ref{eq_dos}) and (\ref{eq_m}),
\begin{eqnarray}
\!\!
\chi =
 -\frac{g_vg_s}{2\pi^2\gamma^2 W} \!
\int_{-\infty}^{\infty} \!\!\! d\vare f(\vare) \!
\int_{-\infty}^{\vare} \!\!\! d\vare'
{\rm Im} \!
\left.
\frac{\partial^2 \Sigma(\vare',B)}{\partial B^2}
\right|_{B=0} \!\! . \enspace
\label{eq_chi_app}
\end{eqnarray}
Let us introduce a variable $X = \vare - \Sigma$
to write $\Sigma$ as a function $(X,B)$ as
\begin{equation}
 \Sigma(\vare,B) \equiv \tilde\Sigma (X,B)= \frac{W(\hbar\omega_B)^2}{2}
\sum_{n=-\infty}^{\infty} \frac{g(\vare_n)}{X-\vare_n}.
\label{eq_XB}
\end{equation}
Using $\partd \Sigma B = - \partd X B$,
the derivative of $\Sigma$ can be written in terms of
those of $\tilde\Sigma$ as
\begin{equation}
 \partd {\Sigma(\vare,B)} B  = 
\left[
1+\partd {\tilde\Sigma(X,B)}X
\right]^{-1}
\partd {\tilde\Sigma(X,B)}B.
\label{eq_dsigma}
\end{equation}
The second-order derivative can be derived similarly as
\begin{eqnarray}
 \partdd \Sigma B &=& 
\left(
1+\partd {\tilde\Sigma}{X}
\right)^{-1}\times \nonumber\\
&&\hspace{-10mm}
\left[
\partdd {\tilde\Sigma}{B}
-
2 \frac{\partial^2\tilde\Sigma}{\partial X \partial B}
\left(
\partd {\Sigma} B
\right)
+
\partdd {\tilde\Sigma}{X}
\left(
\partd {\Sigma} B
\right)^2
\right].
\label{eq_ddsigma}
\end{eqnarray}

Equation (\ref{eq_XB}) can be explicitly written as
\begin{eqnarray}
\tilde\Sigma (X,B)
&=&
\frac{W}{2}
\Delta t
\left[
\frac{h(0)}{2} + 
\sum_{n=1}^{\infty} h(n\Delta t)
\right],
\end{eqnarray}
where $\Delta t = (\hbar\omega_B)^2=2\gamma^2eB/\hbar$, and
$h(x) = 2X g(\sqrt{X})\allowbreak/(X^2-t)$

When $|{\rm Im}X| \gg \hbar\omega_B$, 
$h(t)$ is regarded as smooth with respect to 
increment $\Delta t$,
and we can use an approximation
\begin{eqnarray}
&& \Delta t
\left[
\frac{h(0)}{2} + \sum_{0<n<\infty} h(n\Delta t)
\right]
= \nonumber\\
&&\int_0^\infty h(t) dt - \frac{(\Delta t)^2}{12}
\left[h'(0) + \frac{1}{2}h'(\infty)
\right],
\end{eqnarray}
where $O(\Delta t^3)$ is neglected.
Then we have 
\begin{eqnarray}
 \tilde\Sigma (X,B) - \tilde\Sigma (X,0)
=  -\frac{W}{24}h'(0) (\Delta t)^2 
\end{eqnarray}
which leads to
\begin{eqnarray}
\left.
\frac{\partial \tilde\Sigma}{\partial B} 
\right|_{B=0}
&=& 0 
\label{eq_dsigma2}
\\
\left.
\frac{\partial^2 \tilde\Sigma}{\partial B^2} 
\right|_{B=0}
&=&
-\frac{W}{6}
\left(
\frac{2e\gamma^2}{\hbar}
\right)^2 \frac{1}{X^3}.
\label{eq_ddsigma2}
\end{eqnarray}
With (\ref{eq_chi_app}), (\ref{eq_dsigma}), and (\ref{eq_ddsigma}),
we obtain
\begin{eqnarray}
\chi &=& 
\frac{g_vg_s}{3\pi^2}
\frac{e^2\gamma^2}{\hbar^2}
\int_{-\infty}^{\infty} d\vare f(\vare)
\nonumber\\
\hspace{5mm}
&& \times {\rm Im}
\int_{-\infty}^{\vare}
d\vare'
\left(
 1+\partd {\tilde\Sigma '}{X'}
\right)^{-1}
\left.
 \frac 1 {{X'}^3} 
\right|_{B=0}, \quad
%
\label{eq_chi_app1}
\end{eqnarray}
where $\tilde\Sigma '$ and $X'$ are functions of $\vare'$.
Integration in $\vare'$ can be executed 
with the aid of
\begin{equation}
 d\vare' = 
\left(
 1 + \frac{\partial \tilde\Sigma '}{\partial X'}
\right)
dX',
\end{equation}
and finally obtain 
\begin{eqnarray}
 \chi = 
-\frac{g_vg_s}{6\pi^2}
\frac{e^2\gamma^2}{\hbar^2}
\int_{-\infty}^{\infty} d\vare f(\vare)
{\rm Im}
\left.
 \frac 1 {X^2} 
\right|_{B=0},
\label{eq_chi_app2}
\end{eqnarray}
which is (\ref{eq_chi}).
We can derive the identical equation
starting from the general formula based
on the linear response theory \cite{Fuku}.

We can show that the susceptibility in the present system
has a `sum rule', where
the integration of $\chi(\mu)$ over $\mu$
is a constant independent of the disorder strength.
From (\ref{eq_chi_app2}), we have
\begin{eqnarray}
\int_{-\infty}^\infty \chi(T, \mu)  d\mu
&=& \int_{-\infty}^\infty \chi(0, \vare)d\vare \nonumber\\
\hspace{-10mm}
&=&
-\frac{g_vg_s}{6\pi^2}
\frac{e^2\gamma^2}{\hbar^2} 
\int_{-\infty}^\infty d\vare
\int_{-\infty}^{\vare} d\vare' 
{\rm Im}
\left.
 \frac 1 {{X'}^2} 
\right|_{B=0} 
\nonumber \\
&=&
-\frac{g_vg_s}{6\pi^2}
\frac{e^2\gamma^2}{\hbar^2} 
\int_{-\infty}^\infty d\vare \,
\vare \,
{\rm Im}
\left.
 \frac 1 {X^2} 
\right|_{B=0}.
\end{eqnarray}
By replacing the integrating variable $\vare$ with $X$,
this becomes
\begin{eqnarray}
\int_{-\infty}^{\infty} \chi d\mu 
&=&
-\frac{g_vg_s}{6\pi^2}
\frac{e^2\gamma^2}{\hbar^2} 
\frac{1}{2i}
\oint_C
dX \left(1 + \partd {\tilde\Sigma} X \right)
\frac{\tilde\Sigma + X}{X^2}
\nonumber\\
&=& 
- \frac{g_vg_s}{6\pi^2}\frac{e^2\gamma^2}{\hbar^2}
\frac{1}{2i}
\oint_C \frac{1}{X}
= - \frac{g_vg_s}{6\pi}\frac{e^2\gamma^2}{\hbar^2}
\label{eq_sum_rule}
\end{eqnarray}
where integration path $C$ is a circle with an infinite radius
with clockwise direction,
and we used $\tilde\Sigma \sim O(1/X)$ for large $|X|$.


The susceptibility for the long-ranged disorder (\ref{eq_chi_long})
can be derived in a similar way to the short-ranged case,
while the procedure is rather complicated.
From (\ref{eq_dos_long}) and (\ref{eq_m}),
we obtain
\begin{eqnarray}
\chi &=&
 -\frac{g_vg_s}{2\pi^2\gamma^2 W}
\int_{-\infty}^{\infty} d\vare f(\vare)
\int_{-\infty}^{\vare} d\vare'
\nonumber\\
&&
{\rm Im}
\left.
\frac 1 2
\frac{\partial^2}{\partial B^2}
\left(
\Sigma^+(\vare',B) + \Sigma^-(\vare',B)
\right)
\right|_{B=0}.
\label{eq_chi_app_long}
\end{eqnarray}
We introduce a variable $X^\pm = \vare - \Sigma^\pm$
and define $\Sigma^\pm \equiv \tilde\Sigma^\pm(X^+,X^-,B)$, with
\begin{eqnarray}
 \tilde\Sigma^+ &\equiv& 
W(\hbar\omega_B)^2
\sum_{n=0}^{\infty} 
\frac{X^-g(\vare_n)}
{X^+X^- -\vare_n^2}
\\
\tilde\Sigma^-  &\equiv& 
W(\hbar\omega_B)^2
\sum_{n=1}^{\infty} 
\frac{X^+g(\vare_n)}
{X^+X^- -\vare_n^2}.
\end{eqnarray}
The derivatives of $\Sigma$ can be written in terms of
$\tilde\Sigma$ as
\begin{equation}
 \partd {\Sigma^i} B  = 
A_{ij}\partd {\tilde\Sigma^j}B,
\label{eq_dsigma_long}
\end{equation}
and
\begin{equation}
\partdd {\Sigma^i} B  
= 
A_{ij}
\Big(
\partdd {\tilde\Sigma^j}{B}
-
2 \frac{\partial^2\tilde\Sigma^j}{\partial X^k \partial B}
\partd {\Sigma^k} B
+ \frac{\partial^2\tilde\Sigma^j}{\partial X^k \partial X^l}
\partd {\Sigma^k} B
\partd {\Sigma^l} B
\Big),
\label{eq_ddsigma_long}
\end{equation}
where $i,j,k,l = \pm$,
repeated indices indicate summation, and 
the matrix $A$ is defined as
\begin{equation}
 (A^{-1})_{ij} \equiv 
\delta_{ij} + \partd {\tilde\Sigma^i}{X^j}.
\end{equation}

We can calculate the derivatives of $\tilde\Sigma^\pm$ 
at $B=0$ in a similar way to the short-ranged case,
and then obtain those for $\Sigma^\pm$
through Eqs.\ (\ref{eq_dsigma_long}) and (\ref{eq_ddsigma_long}).
As a result, we have
\begin{eqnarray}
\left.
\partdd {} B
(\Sigma^+ + \Sigma^-)
\right|_{B=0}
&=& 
\frac{1}{1+\alpha+2\beta}
\left(
\frac{2e\gamma^2}{\hbar}
\right)^2
\frac{1}{X^3}
\nonumber\\
&&\hspace{-30.0mm}
\times \Big(
-\frac W 6
+ \frac 1 {1-\alpha} 
\frac {W^2}2
- \frac{2\beta}{(1-\alpha)^2}
\frac{W^2}{4}
\Big) ,
\label{eq_sigpm}
\end{eqnarray}
where 
$ X \equiv \lim_{B\rightarrow 0} X^+ = \lim_{B\rightarrow 0} X^-$,
and
\begin{eqnarray}
 \alpha &=& 2W \int_0^\infty tdt
\frac{g(t)}{X^2-t^2}, 
\label{eq_alpha}
\\
 \beta &=& 2W \int_0^\infty tdt
\frac{-X^2g(t)}{(X^2-t^2)^2}.
\label{eq_beta}
\end{eqnarray}
Substituting (\ref{eq_sigpm}) in (\ref{eq_chi_app_long}), this becomes
\begin{eqnarray}
&& \chi =
-\frac{g_vg_s}{6\pi^2}
\frac{e^2\gamma^2}{\hbar^2}
\int_{-\infty}^{\infty} d\vare f(\vare)
\int_{X(-\infty)}^{X(\vare)} dX'
\nonumber\\
&&
\times {\rm Im}
\frac{1}{{X'}^3}
\left[
- 1+ 3W
\left(
\frac{1}{1-\alpha'} 
- \frac{\beta'}{(1-\alpha')^2} 
\right) 
\right],
\qquad
\label{eq_chi_app_long1}
\end{eqnarray}
where $\alpha'$ and $\beta'$
have the argument $X'$ for $X$ 
in (\ref{eq_alpha}) and (\ref{eq_beta}),
and the integration in $\vare'$ has been replaced by
$d\vare' = (1+\alpha' +\beta') dX$.

In the region $|\vare| \ll \vare_c$,
(\ref{eq_alpha}) and (\ref{eq_beta})
can be approximately written as
$\alpha \approx -W\log(-\vare_c^2/X^2)$
and $\beta \approx W$.
By substituting them in (\ref{eq_chi_app_long1}),
we can execute the integration in $X'$ to obtain
Eq.\ (\ref{eq_chi_long}).
Here the expression of the integrand is valid
only for $|\vare| \ll \vare_c$
while the integration runs over all $\vare$,
but this is justified since the integral converges.


\end{document}